# DAMQ-Based Schemes for Efficiently Using the Buffer Spaces of a NoC Router

Mohammad Ali Jabraeil Jamali, Ahmad Khademzadeh

Department of Computer Science, Islamic Azad University, Science & Research Branch, Tehran, Iran
*m_jamali@itrc.ac.ir* , *zadeh@itrc.ac.ir*

**Abstract**
In this paper we present high performance dynamically allocated multi-queue (DAMQ) buffer schemes for fault tolerance systems on chip applications that require an interconnection network. Two or four virtual channels shared the same buffer space. On the message switching layer, we make improvement to boost system performance when there are faults involved in the components communication. The proposed schemes are when a node or a physical channel is deemed as faulty, the previous hop node will terminate the buffer occupancy of messages destined to the failed link. The buffer usage decisions are made at switching layer without interactions with higher abstract layer, thus buffer space will be released to messages destined to other healthy nodes quickly. Therefore, the buffer space will be efficiently used in case fault occurs at some nodes.

## 1. Introduction

Future system-on-chip (SoC) designs require predictable, scalable and reusable on-chip interconnect architecture to increase reliability and productivity. Current bus-based interconnect architectures are inherently non-scalable, less adaptable for reuse and their reliability decreases with system size. To overcome these problems, it has been proposed to build a message passing network for on-chip communication - network-on-chip (NoC).

Due to the constraints of being in a single chip, using an interconnection network on chip needs be restricted in terms of area. Thus, it is extremely important to design the schemes that require less hardware resources and still provide a good performance. Virtual channel multiplexing across a physical channel is extensively used to boost performance and avoid deadlock.

As virtual channels are not equally used in many applications, if they share a common buffer, the whole buffer space will be better utilized. In this paper we present schemes that are based on a dynamically allocated multi-queue (DAMQ) buffer. These schemes provide similar performance as other statically allocated multiple-queue (SAMQ) buffers using less hardware and, therefore, requiring less hardware.

In order to improve the reliability of SoCs, their interconnect infrastructures must be designed such that fabrication and life-time faults can be tolerated. These irrecoverable faults influence the behavior of NoC fabrics and consequently degrade the system performance. Therefore, achieving on-chip fault tolerant communication is becoming increasingly important in presence of such permanent faults. A fault tolerant algorithm distinguishes from deterministic one according to the fact that it can provide an alternative path so that the message wouldn't be blocked by a faulty component [1], [2]. Consequently we investigate the applicability of partially adaptive algorithms to achieve a certain degree of fault tolerance in NoC communication fabrics.

The paper is organized as follows. In section 2, we review the fault tolerant odd-even algorithm in mesh-based NoC while in section 3, the DAMQ shared (DAMQS) and DAMQ all shared (DAMQAS) buffer schemes are discussed. Experimental results, that show the performance of the proposed approaches, are presented in Section 4 followed by conclusions in Section 5.

## 2. Fault Tolerant Odd-even Algorithm in Mesh-Based NoC

Fault tolerance is the ability of the network to function in the presence of component failure. The turn model is a well known partial adaptive routing algorithm, widely investigated for multi-processor environments [3]. The odd-even turn model facilitates deadlock-free routing in mesh network of a NoC.

The detail of odd-even turn model is explained with the help of Fig. 1. In the figure, the odd-even routing algorithm has been illustrated.

IJCSI

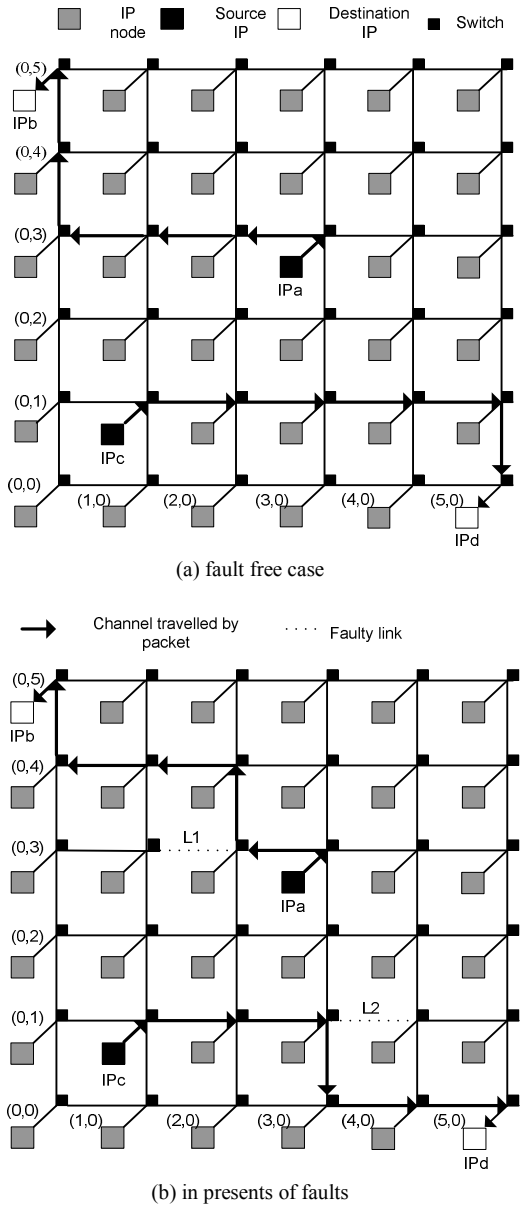

Fig. 1  A sample for Odd-even turn model algorithm.

In Fig. 1. (a), source IPa is trying to send a packet to destination IPb. Two situations are presented in Fig. 1. (a) and (b) respectively. When the network is fault-free, the packet is first routed in x direction before being routed to y direction. When one link L1 is faulty in the path, if normal deterministic x-y routing was adopted, there is no path for the packet to move towards the destination. For odd-even turn algorithm, the packet is routed one hop perpendicular to the top, then toward the destination. As shown in Fig. 1. (b), suppose source IPc is trying to communicate to the destination IPd, and the link L2 is faulty. In this case the packet is routed, one hop toward the down, then toward the destination.

## 3. DAMQS and DAMQAS Buffer Schemes

DAMQ dynamically allocates buffer blocks according to the packet received. Compared with SAMQ, the advantage of DAMQ is that it uses efficiently the buffer space by applying free space to any incoming packet regardless its destination output port.

DAMQ All (DAMQA) is based on self-compacting buffer scheme. Two buffer slots are reserved for each virtual channel before the buffer accepts any incoming flit and during buffers operation [4] [5] [6].

DAMQS buffer combines the buffer for virtual channels from two different physical channels. We combine the buffer space for east X and south Y virtual channels to build one physical buffer, and west X and north Y forms another buffer group. There are two buffers for four physical channels; each buffer is shared by eight virtual channels, and has two read ports and write ports respectively.

As shown in Fig. 2, in this way, the shared space is placed in the middle of the two buffer regions of two virtual link groups then the two buffer regions expand towards center of the free buffer space. This way, there will be less data shifts when a flit is saved into buffer because the movement of one region doesn't depend on shift of another group.

When a node or a physical channel is deemed as faulty, the previous hop node will terminate the buffer occupancy of messages destined to the failed link. Therefore, the buffer space will be efficiently used in case fault occurs at some nodes. For example, if node X is connecting to node Y and Z and there are message flows from X to Y and X to Z. When Z fails, there are probably still a number of message flits left in X's buffer. It will improve the system performance if the buffer space occupied by Z's message can be allocated to Y's message quickly.

DAMQAS buffer combines the buffer for virtual channels from four different physical channels. There is one buffer for four physical channels; the buffer is shared by sixteen virtual channels, and has four read ports and write ports respectively.

As shown in Fig. 2. (a,c), two buffer slots are reserved for each virtual channel before any flit comes in the buffer. Virtual channels on X dimension start to occupy the buffer from lower end of the whole buffer space while virtual channels on Y dimension start using buffer on another end. The buffer space of groups expands and compresses in opposite direction. When a buffer group accepts a flit it expands, when it dispatches a flit it compresses.

The reserve space (RS) is always kept if there is no flit or only one flit in the buffer region for a specific virtual channel [4].

As shown in Fig. 2. (b,d), when the buffer performs shift up or shift down operations, the RSs are also shifted. When a virtual channel accepts a flit, it first uses its RS. If RS is used up, buffer space of the whole group expands toward another group's buffer space to produce a slot. At any time, the number of current flits in buffer plus the number of reserved slots equals to the total amount of buffer slots. Therefore, one or more virtual channels which have the flits come into the buffer at earlier time can never deprive the chance for other virtual channels which get flits later than them to get buffer.

Also, in case the earlier coming packets are blocked in the buffer, since there is still reserved space for other virtual channels, the network traffic will keep flowing; therefore the performance of the switch is enhanced. Moreover, as virtual channels from two or four physical channels are sharing the buffer, the buffer space is more efficiently used by the incoming flits. The results will be shown in next section.

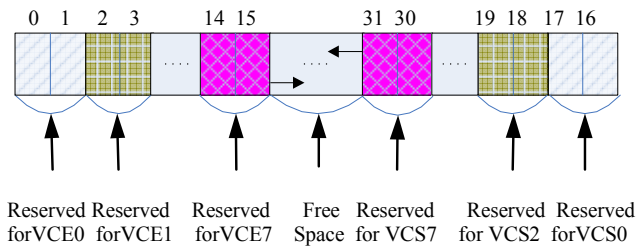

(a) DAMQS at initial state (# VC=8).

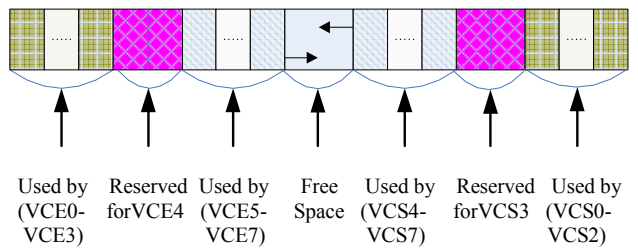

(b) DAMQS at operation state (# VC=8).

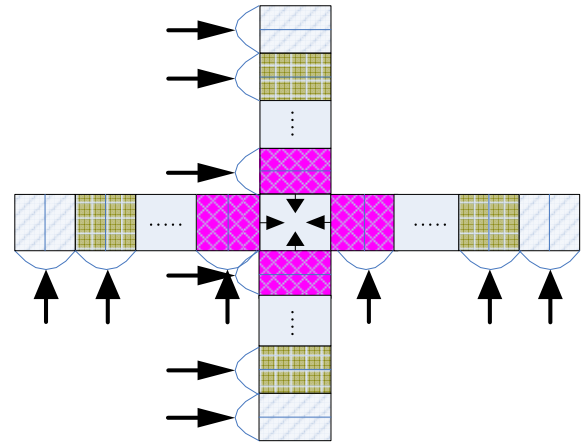

(c) DAMQAS at initial state (# VC=8).

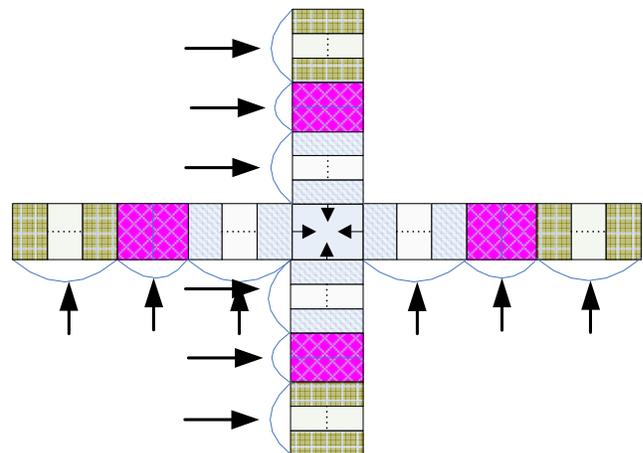

(d) DAMQAS at operation state (# VC=8).

Fig. 2.   DAMQS and DAMQAS buffer spaces status. VCE (VC East), VCS (VC South)

## 4. Experimental Results

In this section, we consider a system consisting of 64 IP blocks mapped onto mesh-based NoC architecture as shown in Fig. 1. Packets size is set to 32 flits. Virtual channels number for each physical channel is 4. Faults are generated randomly in the network. We set the buffer size (BS) for each virtual channel to 4 flits when DAMQA and SAMQ are used. Since four virtual channels are multiplexing cross one physical channel, the buffer size for each direction of a duplex physical channel is 16 flits when these two buffer schemes are evaluated. To examine the performance of DAMQS and DAMQAS with regard to the relationship between buffer size and network performance, we use flits buffer with different sizes. Since the network performance is greatly influenced by the traffic pattern, we applied two different traffic patterns, including synthetic traffic pattern (uniform) and a real-life traffic pattern (telecom) is retrieved from "E3S" benchmark suite [7], which contain 30 tasks. Fig. 3 (a, b) shows the message latency as function of injection load with 2% fault rate (fr) for two uniform and telecom traffic patterns, respectively. When we further increase the traffic load and the fault rate after the network starts getting saturated, DAMQAS shows higher latency than other schemes. The reason is DAMQAS holds much more flits in the buffer than other schemes.

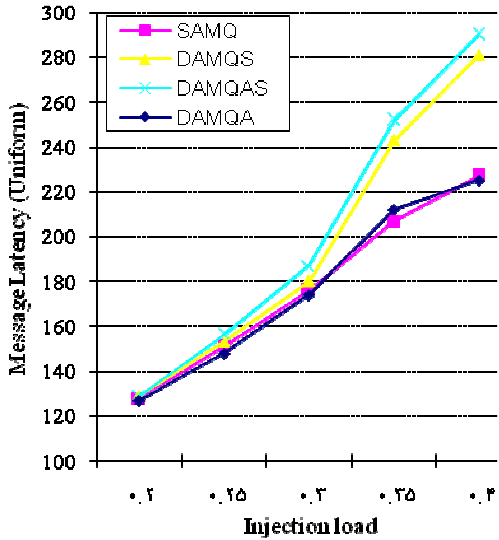

(a) Uniform traffic

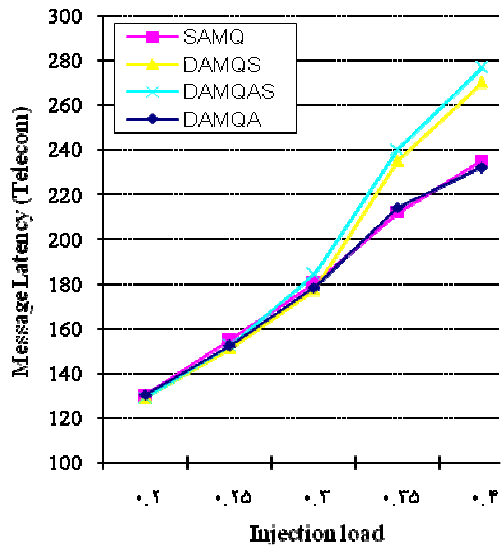

(b) Telecom traffic

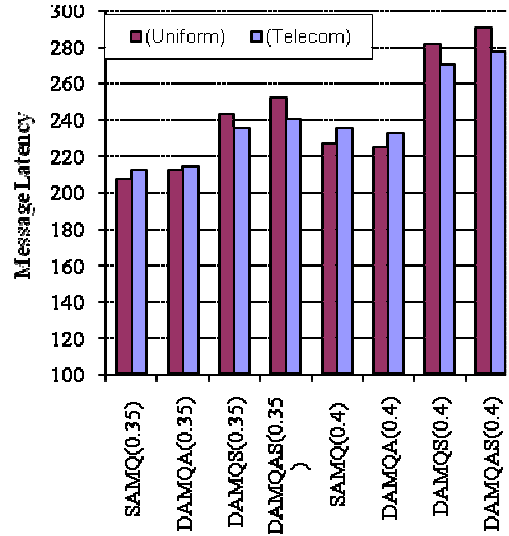

(c) Latency Comparison

Fig. 3. Message latency (2% fr)

In Fig. 3 (c), we compare the message latency of network for two different traffic patterns under different injection rates.

Fig. 4 (a, b) shows the throughput characteristics of the NoC for each buffer scheme in presence of permanent faults with different fault rates for two uniform and telecom traffic patterns, respectively.

As shown in Fig. 4, DAMQAS and DAMQS can provide a better performance in presence of faults than other schemes. For example in uniform traffic, 14-flit DAMQS with 0% fr and 8-flit DAMQS with 4% fr achieves approximately the same maximum throughput as a 16-flit DAMQA as shown in Fig. 4 (b). Also, a 13-flit DAMQAS with 0% fr and 6-flit DAMQAS with 4% fr achieves approximately the same maximum throughput as 16-flit DAMQA. This is to say, to provide a similar network performance on very limited buffer resource, DAMQA with 0% fr achieves similar throughput with 12.5% and 18.75% more buffer space than DAMQS and DAMQAS with 0% fr, respectively and DAMQA with 4% fr achieves similar throughput with 50% and 62.5% more buffer space than DAMQS and DAMQAS with 4% fr, respectively. For telecom traffic, 13-flit DAMQS with 0% fr and 7-flit DAMQS with 4% fr achieves approximately the same maximum throughput as a 16-flit DAMQA as shown in Fig. 4 (b). Also, a 12-flit DAMQAS with 0% fr and 5-flit DAMQAS with 4% fr achieves approximately the same maximum throughput as 16-flit DAMQA. This result shows that, DAMQA with 0% fr achieves similar throughput with 18.75% and 25% more buffer space than DAMQS and DAMQAS with 0% fr, respectively and DAMQA with 4% fr achieves similar throughput with 56.25% and 68.75% more buffer space than DAMQS and DAMQAS with 4% fr, respectively.

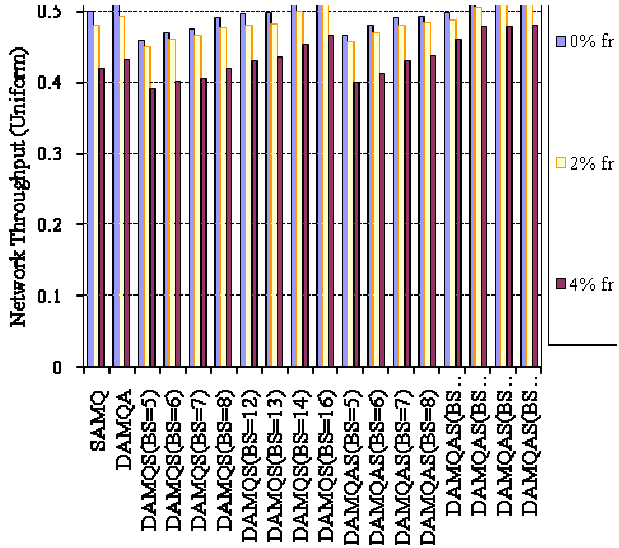

(a) Uniform traffic

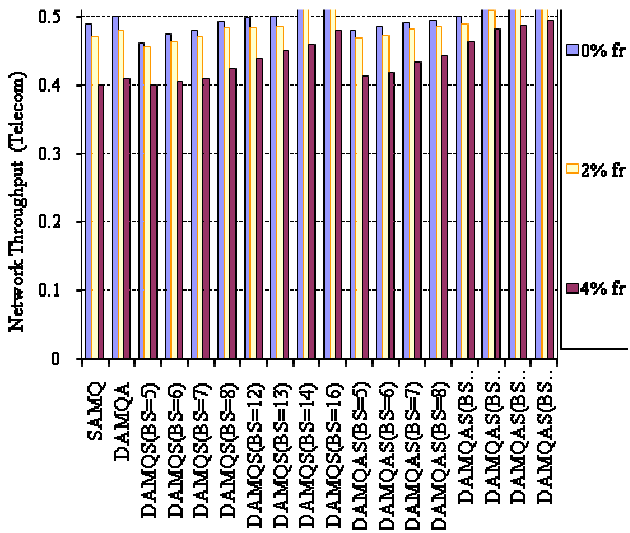

(b) Telecom traffic
Fig. 4. Network throughput

The total buffer space (BUF$_{total}$) can be obtained by the following formula:
$$BUF_{total} = (N \times PHY - 4\sqrt{N}) \times VC \times VB$$

where PHY is the physical channel corresponding to a node port, N is the count of nodes, VC is the count of virtual channel multiplexing a physical channel and VB is the buffer size of a virtual channel. The total available buffer space is 3584 flits in our simulations. We considered high injection load of 0.35. As shown in Fig. 5, DAMQS and DAMQAS can provide a better performance than DAMQA and SAMQ.

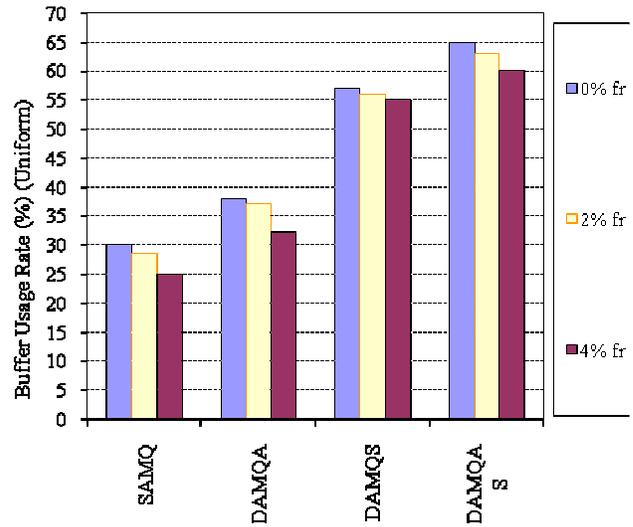

(a) Uniform traffic

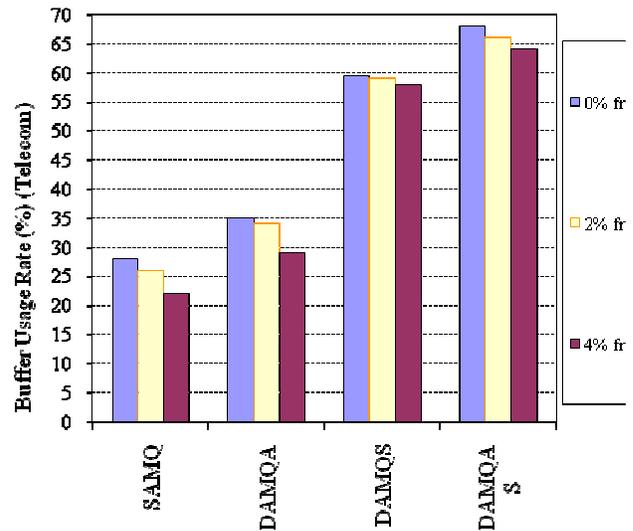

(b) Telecom traffic
Fig. 4. Buffer usage rate

Comparing uniform traffic results with telecom traffic results, it is clear that the DAMQS and DAMQAS are even more beneficial for real benchmarks. Indeed, compared to the uniform traffic pattern in Fig. 3 (a), Fig. 4 (a) and Fig. 5 (a), the traffic patterns for real benchmarks are much more unbalanced (for instance, some of the channels have even a load of zero), which makes the idea of share buffer more attractive.

## 5. Conclusions

For deep sub-micron VLSI processes, the life-time reliability of devices is likely to be compromised by effects such as electromigration and material ageing. Consequently, the performance of NoC interconnect architectures will be severely affected due to presence of permanent faults. Though deterministic routing is very easy to implement, it fails to sustain the desired level of performance in presence of permanent faults. When an adaptive routing protocol such as odd-even turn algorithm is used for the NoC, DAMQAS and DAMQS are excellent schemes to optimize buffer management providing a good throughput when the network has a larger load in presence of faults. They can utilize significantly less buffer space with further increase of the fault rate without sacrificing the network performance. As shown in our simulation results, these buffer schemes can provide marginally higher throughput than traditional SAMQ when same amount of resource is used, this is due to the fact that buffer cannot play a major role in determining the network performance in terms of throughput or latency. However, the results show that these schemes can use significantly less hardware to provide a same performance as traditional SAMQ buffer. The simulation results show that the proposed DAMQ-Based schemes are indeed beneficial for real-world applications characterized by real-life traffic patterns.

**Mohammad Ali Jabraeil Jamali.** received his B.Sc. in Electrical Engineering and M.Sc. in Computer Hardware Engineering from Oromie University in 1994 and Tehran Science and Research Branch of Islamic Azad University in 2003, respectively. Currently, he is a Ph.D. student and Faculty Member at the Tehran Science and Research Branch of Islamic Azad University under supervision of Dr. Ahmad khademzadeh and Islamic Azad University, Shabestar Branch, respectively. His research interests include VLSI Design, Interconnection Network, Fault Tolerant, Ad hoc Network, Sensor Network and Computer Architectures.

**Ahmad Khademzadeh**. is currently associate professor in Iran Telecommunication Research Center. His research interests include VLSI Design, Interconnection Network, Fault Tolerant and Computer Architectures.